\documentclass[referee]{aa}  

\usepackage{lineno}
\usepackage{graphicx}
\usepackage{txfonts}

\allowdisplaybreaks

\renewcommand{\Vec}[1]{\mbox{\boldmath$#1$}}

\newcommand*\ii{ \dot{\mbox{\i\!\!\i}} }

\begin{document}

\title{Non-singular recursion formulas for third-body perturbations in mean vectorial elements}
\titlerunning{Third-body perturbations in mean vectorial elements}
\authorrunning{M. Lara et al.}
%\subtitle{I. Overviewing the $\kappa$-mechanism}

\author{M. Lara\inst{1,3}
        \and
				A.~J. Rosengren\inst{2}
				\and
        E. Fantino\inst{3}}

   \institute{Scientific Computing Group, University of La Rioja, Calle Madre de Dios 53, 26006 Logroño, Spain
	 %\email{martin.lara@unirioja.es}
		\and{Department of Aerospace \& Mechanical Engineering, The University of Arizona, 1130 N. Mountain Ave., P.O. Box 210119, Tucson, Arizona}
        \and
         Department of Aerospace Engineering, Khalifa University of Science and Technology, P.O. Box 127788, Abu Dhabi, United Arab Emirates\\
				\email{martin.lara@unirioja.es; elena.fantino@ku.ac.ae}}

   \date{Received XXX; accepted YYY}

\abstract{The description of the long-term dynamics of highly elliptic orbits under third-body perturbations may require an expansion of the disturbing function in series of the semi-major axes ratio up to higher orders. To avoid dealing with long series in trigonometric functions, we refer the motion to the apsidal frame and efficiently remove the short-period effects of this expansion in vectorial form up to an arbitrary order. We then provide the variation equations of the two fundamental vectors of the Keplerian motion by analogous vectorial recurrences, which are free from singularities and take a compact form useful for the numerical propagation of the flow in mean elements.}

\keywords{Celestial Mechanics}

   \maketitle
\section{Introduction}

Perturbed Keplerian motion is a multi-scale problem, in which the orbital elements evolve slowly when compared to the change with time of ephemeris, whose fast evolution is determined by the rate of variation of the mean anomaly. Usual integration schemes that look for the separation of fast and slow frequencies of the motion may be superior to the simpler integration with the Cowell method, in which the Newtonian acceleration is directly integrated in rectangular coordinates, in particular problems, and formulations based on Hansen's ideal frame concept \citep{Hansen1857} closely approach to this decoupling (see \cite{Lara2017if} and references therein). Also, the use of symplectic integrators is widely adopted in the case of Hamiltonian perturbation problems \citep{LaskarRobutel2001,Blanesetal2013}.
\par

On the other hand, the effective decoupling of short- and long-period effects is achieved with perturbation methods. After removing short-period effects in an averaging process, the long-term behavior of the (mean) orbital elements is efficiently integrated from the variation of parameters equations. The parameters, which are constant in the pure Keplerian motion, can take different representations, as classical Keplerian elements or non-singular variables (see \cite{Hintz2008} for a survey), but are always related to the two fundamental vectors of the orbital motion. Namely, the angular momentum vector, which determines the orientation of the orbital plane, and the eccentricity vector, which determines the shape and orientation of the reference ellipse in that plane---the latter being a non-dimensional version of the classical Laplace vector, which has dimensions of the gravitational parameter, or of the Runge-Lenz vector that has dimensions of angular momentum. While vectorial formulations normally introduce redundancy by increasing the dimension of the differential system, they commonly admit a compact and symmetric formulation of the right side of the variation equations that renders faster evaluation \citep{Allan1962,Musen1963JGR,RoyMoran1973},\footnote{A brief historical review on the topic can be consulted in \cite{Deprit1975}, with additional details in \cite{RosengrenScheeres2014}.} and may even disclose integrability \citep{Deprit1984} (see also \cite{MignardHenon1984,RichterKeller1995}).
\par

The vectorial formulation has shown to be useful in the case of third-body perturbations \citep{BreiterRatajczak2005,Correiaetal2011,KatzDongMalhotra2011}. In particular, it may be an efficient alternative to classical formulations when the orbits are highly elliptic, which is a common case in extrasolar planetary systems \citep{LeePeale2003,MigaszewskiGozdziewski2008}, in artificial satellite theory \citep{LaraSanJuanLopezCefola2012,LaraSanJuanHautesserres2018}, and in hierarchical $n$-body systems in general \citep{Hamersetal2015,Will2017}. These kinds of orbits may need higher degrees in the expansion of the third-body disturbing function to provide a reasonable approximation of the dynamics even in the simplifications offered by the secular dynamics \citep{BeaugeMichtchenko2003,LibertSansottera2013,Andrade-Inesetal2016,Andrade-InesRobutel2018,SansotteraLibert2019}. %[{\color{magenta}preprint $G^3$ needed}], 
This fact has motivated the recent appearance of general expressions of the expansion of the disturbing function in powers of the ratio of the disturbing and disturbed semi-major axes, including the secular terms \citep{LaskarBoue2010,Mardling2013,PalacianVanegasYanguas2017}. These expressions rely on classical elements of the Keplerian motion and apply for any eccentricity and inclination.
\par

We constrain ourselves to the restricted approximation, in which the mass of the lighter body does not affect the motion of the primaries. This is a reasonable approximation in artificial satellite theory as well as in some problems of the Solar System dynamics, yet it does not apply to the dynamics of extrasolar planetary system. For the restricted case, we make use of the apsidal frame formulation, which effectively displays the fast and slow components of the orbital motion, and use perturbation theory to remove the short-period components of the third-body disturbing function. The third-body direction is assumed to be a known function of time, given by an ephemeris, but we avoid time related issues in the averaging by constraining ourselves to the usual case in which the position of the third body can be taken as fixed during one orbital period. Instead of relying on the classical Hansen expansions or related eccentricity functions  \cite{Hansen1855,Kaula1962,Giacaglia1974,Lane1989,Cellettietal2017}, we average the Legendre polynomials expansion of the third-body potential in closed form of the eccentricity after the usual reformulation in terms of the eccentric anomaly \citep{Deprit1983,Kelly1989}.
\par

The decoupling of short-period effects is achieved after the standard Delaunay normalization \citep{Deprit1982}, which we extend only to the first order in the Hamiltonian perturbation approach; that is, we do not consider the possible coupling between the different terms in which the third-body perturbation is expanded. However, while the normalization is properly carried out in Delaunay canonical variables, which are the action-angle variables of the Kepler problem, the canonical variables related to the Keplerian elements are collected in vectorial form along the whole normalization procedure. In this way we obtain alternative, general expressions for the expansion of the third-body disturbing function in mean vectorial elements up to an arbitrary degree. For completeness,  the generating function of the infinitesimal contact transformation from mean to osculating elements is also provided in vectorial elements up to an arbitrary degree of the expansion.
\par

The Hamilton equations of the mean elements Hamiltonian are then computed to obtain analogous general expression for the variation equations of the flow in Delaunay (mean) elements, which, as expected, are flawed by the appearance of the eccentricity and the sine of the inclination as divisors. Singularities are avoided reformulating the flow in different sets of canonical and non-canonical variables. In particular, we provide vectorial, non-singular expressions for the variation equations of the eccentricity and angular momentum vectors in mean elements. These expressions can be used to extend to an arbitrary degree existing lower-order truncations in the literature \citep{Allan1962,KatzDongMalhotra2011}.
\par

\section{Third-body disturbing potential}

Let $(O,\Vec{i},\Vec{j},\Vec{k})$ be an orthonormal frame with origin $O$ in the center of mass of a central attracting body and fixed directions defined by the unit vectors $\Vec{i},\Vec{j},\Vec{k}$. Let $\Vec{r}$ and $\Vec{r}_\star$ be the position vectors of a massless and a massive body, respectively, of modulus $r=\|\Vec{r}\|$ and $r_\star=\|\Vec{r}_\star\|$. When ${r}\ll{r}_\star$, the third-body disturbing potential
\begin{equation}
\mathcal{V}_\star=-\frac{\mu_\star}{r_\star}\left(\frac{r_\star}{\|\Vec{r}-\Vec{r}_\star\|}-\frac{\Vec{r}\cdot\Vec{r}_\star}{r_\star^2}\right),
\end{equation}
where $\mu_\star$ is the third-body gravitational parameter, is customarily replaced by the Legendre polynomials expansion
\begin{equation} \label{thirdbody}
\mathcal{V}_\star=-\frac{n_\star^2a_\star^3}{r_\star}\chi_\star\sum_{i\ge2}\frac{r^i}{r_\star^i}P_i(\cos\psi_\star),
\end{equation}
in which $\mu_\star$ has been rewritten in terms of the mean motion $n_\star$ and the semimajor axis $a_\star$ of the third body orbit relative to the central body, $\chi_\star$ is the mass ratio of the system, $\cos\psi_\star=\hat{\Vec{r}}\cdot\hat{\Vec{r}}_\star$, with $\hat{\Vec{r}}=\Vec{r}/r$ and $\hat{\Vec{r}}_\star=\Vec{r}_\star/r_\star$, and the Legendre polynomials $P_i$ are given by  the usual expansion of Rodrigues' formula, namely
\begin{equation} \label{LegendreP}
P_i(\cos\psi_\star) = \frac{1}{2^i}\sum_{l=0}^{\lfloor{i/2}\rfloor}(-1)^l\binom{i}{l}\binom{2i-2l}{i}\cos^{i-2l}\psi_\star,
\end{equation}
with $\lfloor{i/2}\rfloor$ denoting the integer part of the division $i/2$. Recall that, as usual, the term $-\mu_\star/r_\star$ is neglected in Eq.~(\ref{thirdbody}) because it brings null contribution to the disturbing acceleration of the massless body.
\par

Define now the apsidal, moving frame $(O,\hat{\Vec{e}},\Vec{b},\Vec{n})$ with the same origin as before and the unit vectors $\Vec{n}$ in the direction of the (instantaneous) angular momentum vector (per unit of mass)
\begin{equation} 
\Vec{G}=G\Vec{n}=\Vec{r}\times\Vec{v},
\end{equation} 
where $G=\|\Vec{G}\|$ and $\Vec{v}=\mathrm{d}\Vec{r}/\mathrm{d}t$, with $t$ denoting time. The unit vector $\hat{\Vec{e}}$ has the direction of the (instantaneous) eccentricity vector
\begin{equation} 
\Vec{e}=e\hat{\Vec{e}}=\frac{\Vec{v}\times\Vec{G}}{\mu}-\frac{\Vec{r}}{r},
\end{equation} 
where $e=\|\Vec{e}\|$ is the (instantaneous) eccentricity of the orbit of the massless body, and $\mu$ is the central body's gravitational parameter. Finally, the unit vector
\begin{equation} 
\Vec{b}=\Vec{n}\times\hat{\Vec{e}},
\end{equation} 
defines the binormal direction, in this way completing a direct orthonormal frame.
\par

For convenience, we define the unit vector $\Vec\ell$ in the direction of the ascending node, given by
\begin{equation} \label{vl}
\Vec{k}\times\Vec{n}=\Vec\ell\sin{I},
\end{equation}
where
\begin{equation}
I=\arccos(\Vec{k}\cdot\Vec{n}),
\end{equation}
is the inclination angle of the (instantaneous) orbital plane with respect to the $(\Vec{i},\Vec{j})$ plane. Besides,
\begin{equation} \label{swcw}
\Vec\ell\cdot\hat{\Vec{e}}=\cos\omega, \qquad
\Vec\ell\cdot\Vec{b}=-\sin\omega, 
\end{equation}
where $\omega$ is the argument of the pericenter. Then, it is simple to check that
\begin{eqnarray} \label{eklb}
\hat{\Vec{e}}\cdot\Vec{k} &=& \sin{I}\sin\omega=-\Vec\ell\cdot\Vec{b}\sin{I}, \\ \label{bkle}
\Vec{b}\cdot\Vec{k} &=& \sin{I}\cos\omega=\Vec\ell\cdot\hat{\Vec{e}}\sin{I}.
\end{eqnarray}
\par

Finally, if we denote with $h$ the longitude of the node, the components of the apsidal frame in the inertial frame are given by the rotations
\begin{equation} \label{rotate}
\left(\hat{\Vec{e}},\Vec{b},\Vec{n}\right)=R_3(-h)\,R_1(-I)\,R_3(-\omega),
\end{equation}
where
\[
R_1(\alpha)=
\left(
\begin{array}{ccc}
 1 & 0 & 0 \\
 0 & \cos\alpha & \sin\alpha \\
 0 & -\sin\alpha & \cos\alpha \\
\end{array}
\right),
\qquad
R_3(\alpha)=
\left(
\begin{array}{ccc}
 \cos\alpha & \sin\alpha & 0 \\
 -\sin\alpha & \cos\alpha & 0 \\
 0 & 0 & 1 \\
\end{array}
\right)
\]
are standard rotation matrices.
\par

When the direction of the massless body is given by its components in the apsidal frame
\begin{equation} 
\hat{\Vec{r}}=\hat{\Vec{e}}\cos{f}+\Vec{b}\sin{f},
\end{equation} 
with $f$ being the true anomaly, we get
\begin{equation} 
\cos\psi_\star=(\hat{\Vec{e}}\cdot\hat{\Vec{r}}_\star)\cos{f}+(\Vec{b}\cdot\hat{\Vec{r}}_\star)\sin{f},
\end{equation} 
which immediately discloses the short-period terms affecting Eq.~(\ref{LegendreP}).
\par

However, the short-period terms of the third-body disturbing potential in Eq.~(\ref{thirdbody}) do not limit to those contributed by the Legendre polynomials, and they are better handled when written in terms of the eccentric anomaly $u$ contrary to the true one. This is done using the geometric relations
\begin{equation} \label{a:f2u}
\sin{f}=\frac{a}{r}\eta\sin{u}, \qquad \cos{f}=\frac{a}{r}(\cos{u}-e),
\end{equation}
where the eccentricity function $\eta=(1-e^2)^{1/2}$ is used for convenience. Recalling, besides, that
\begin{equation} \label{r2u}
r=a(1-e\cos{u}),
\end{equation}
after some rearrangement we get
\begin{equation} \label{V3b}
\mathcal{V}_\star=
-\frac{\mu}{2a}\chi_{\star}\frac{n_\star^2}{n^2}\frac{a_\star^3}{r_\star^3}\frac{a}{r}\sum_{i\ge2}\frac{1}{2^{i-1}}\frac{a^{i-2}}{r_\star^{i-2}}V_i,
\end{equation}
in which the non-dimensional functions $V_i$ take the form
\begin{equation} \label{Vi}
V_i=\sum_{l=0}^{\lfloor{i/2}\rfloor}(-1)^l\binom{i}{l}\binom{2i-2l}{i}
\frac{r^{2l+1}}{a^{2l+1}}\left[(\hat{\Vec{e}}\cdot\hat{\Vec{r}}_\star)(\cos{u}-e)+(\Vec{b}\cdot\hat{\Vec{r}}_\star)\eta\sin{u}\right]^{i-2l}.
\end{equation}
The reasons for the particular arrangement of Eq.~(\ref{V3b}), in which we left the coefficient $a/r$ out of the summation, is that the removal of short-period terms from $\mathcal{V}_\star$, which will be carried out by Eq.~(\ref{intVlinu}), is easily achieved in closed form of the eccentricity by taking advantage of the differential relation between the mean and true anomalies. 
\par

Next, we substitute Eq.~(\ref{r2u}) into Eq.~(\ref{Vi}) and use the binomial expansion to get
\begin{equation} \label{binria}
\frac{r^{2l+1}}{a^{2l+1}}=(1-e\cos{u})^{2l+1}=\sum_{j=0}^{2l+1}\binom{2l+1}{j}(-1)^je^j\cos^ju.
\end{equation}
Analogously, the binomial expansion is applied to the last term in the right side of Eq.~(\ref{Vi}), to obtain
\begin{eqnarray} \label{binshp}
&& \left[(\hat{\Vec{e}}\cdot\hat{\Vec{r}}_\star)(\cos{u}-e)+(\Vec{b}\cdot\hat{\Vec{r}}_\star)\eta\sin{u}\right]^{i-2l}= \nonumber \\ 
&& \hspace{1cm} \sum_{k=0}^{i-2l}\binom{i-2l}{k}(\hat{\Vec{e}}\cdot\hat{\Vec{r}}_\star)^k(\cos{u}-e)^k
(\Vec{b}\cdot\hat{\Vec{r}}_\star)^{i-2l-k}\eta^{i-2l-k}\sin^{i-2l-k}u,
\end{eqnarray}
where
\begin{equation} \label{bincos}
(\cos{u}-e)^k=\sum_{m=0}^k\binom{k}{m}(-1)^me^m\cos^{k-m}u.
\end{equation}
Plugging Eqs.~(\ref{binria})--(\ref{bincos}) into Eq.~(\ref{Vi}), we finally get
\begin{eqnarray} \label{Viexpanded}
V_i &=& \sum_{l=0}^{\lfloor{i/2}\rfloor}\binom{i}{l}\binom{2i-2l}{i}\sum_{j=0}^{2l+1}\binom{2l+1}{j}\sum_{k=0}^{i-2l}\binom{i-2l}{k}\sum_{m=0}^k\binom{k}{m}e^{j+m} \nonumber\\ 
  & & \times(-1)^{j+l+m}\eta^{i-2l-k}(\hat{\Vec{e}}\cdot\hat{\Vec{r}}_\star)^k(\Vec{b}\cdot\hat{\Vec{r}}_\star)^{i-2l-k}\cos^{j+k-m}u\sin^{i-2l-k}u.
\end{eqnarray}

\section{Short-period averaging}

Since third-body perturbations derive from the potential in Eq.~(\ref{thirdbody}), we can take advantage of the Hamiltonian formalism. Thus
\begin{equation} \label{Hamosc}
\mathcal{H}=-\frac{\mu}{2a}+\mathcal{V}_\star,
\end{equation}
where $\mathcal{H}$ must be expressed in some set of canonical variables. In particular, we use Delaunay variables $(\ell,g,h,L,G,H)$ where the coordinates are the mean anomaly $\ell$, the argument of the periapsis $g$, and the longitude of the node $h$, and their conjugate momenta are the Delaunay action $L=\sqrt{\mu{a}}$, the modulus of the angular momentum vector $G$, and is its projection along the $\Vec{k}$ direction $H$, respectively. A modern derivation of this useful set of variables can be found, for instance, in \cite{LaraTossa2016}. 
\par

Disregarding short-period effects, the orbit evolution is customarily studied in mean elements $(\ell',g',h',L',G',H')$, which aim to represent the average value of the true, osculating variables. The transformation from mean to osculating variables is found using the tools of perturbation theory. To avoid time dependency issues introduced by the third-body ephemeris $\hat{\Vec{r}}_\star\equiv\hat{\Vec{r}}_\star(t)$, we used the extended phase space formulation. In particular, we rely on the Lie transforms method \citep{Hori1966,Deprit1969}, which is considered standard these days and is thoroughly described in the literature (see \cite{MeyerHall1992,BoccalettiPucacco1998v2}, for instance). The details of the transformation will be presented elsewhere, and we focus here on the averaging of the disturbing potential. Still, since its computation is immediate once the disturbing potential has been averaged, we also provide the generating function from which the short-period corrections are derived.
\par

The short-period elimination is carried out in closed form of the eccentricity. It is effectively achieved with the help of the differential relation between the eccentric and true anomalies $\mathrm{d}\ell=(1-e\cos{u})\,\mathrm{d}u$, which is obtained by differentiation of Kepler's equation. Thus, on account of Eq.~(\ref{r2u}), 
\begin{equation} \label{intVlinu}
\langle\mathcal{V}_\star\rangle_\ell=\frac{1}{2\pi}\int_0^{2\pi}\mathcal{V}_\star\mathrm{d}\ell
=\frac{1}{2\pi}\int_0^{2\pi}\mathcal{V}_\star\frac{r}{a}\mathrm{d}u,
\end{equation} 
from which, after replacing Eq.~(\ref{V3b}), we obtain
\begin{equation} \label{H02}
\langle\mathcal{V}_\star\rangle_\ell=-\chi_{\star}\frac{\mu}{2a}\frac{n_\star^2}{n^2}\frac{a_\star^3}{r_\star^3}\sum_{i\ge2}\frac{1}{2^{i-1}}\frac{a^{i-2}}{r_\star^{i-2}}\langle{V}_i\rangle_u,
\end{equation}
where
\begin{equation} \label{Viua}
\langle{V}_i\rangle_u=\frac{1}{2\pi}\int_0^{2\pi}V_i\mathrm{d}u.
\end{equation}
\par

In view of Eq.~(\ref{Viexpanded}), the only terms that remain below the integral symbol of Eq.~(\ref{Viua}) are of the form $\sin^au\cos^bu$ with $a$ and $b$ integers. Terms of this kind are easily integrated when expressed as a Fourier series in $u$. Straightforward computations using standard relations between exponentials and circular functions (see \cite{Kaula1966}, for instance), yield
\begin{eqnarray} \label{fourieru}
\cos^{j+k-m}u\sin^{i-2l-k}u &=& 
\frac{(-\ii)^{i-k-2l}}{2^{i+j-m-2l}}\sum_{q=0}^{i-k-2l}\sum_{t=0}^{j+k-m}\binom{i-k-2l}{q}(-1)^q \nonumber\\ 
&& \times\binom{{j+k-m}}{t}\left[\cos2(\tilde{q}-q)u+\ii\sin2(\tilde{q}-q)u\right],
\end{eqnarray}
where $\ii$ notes the imaginary unit, and we abbreviated
\begin{equation} \label{qtilde}
\tilde{q}=\frac{i+j-m}{2}-l-t.
\end{equation}
\par

Now, it becomes obvious that the only non-periodic terms of Eq.~(\ref{fourieru})
\begin{equation} 
\Xi_{j+k-m}^{i-2l-k}=\left\langle\cos^{j+k-m}u\sin^{i-2l-k}u\right\rangle_u,
\end{equation}
are those such that $q=\tilde{q}$, a condition that is only accomplished if $i+j-m$ is even, which in turn implies that $i+j+m$ is also even. Besides, since Eq.~(\ref{fourieru}) must be free from imaginary terms, it happens that $i-k$, the exponent of the imaginary unit, must be even in the case of non-periodic terms, yielding the numeric coefficient
\begin{equation}
\Xi_{2a}^{2b}=\frac{1}{2^{2a+2b}}\sum_{t=0}^{2a}\binom{2b}{b+a-t}(-1)^{a-t}\binom{2a}{t},
\end{equation}
where $a=\frac{1}{2}(j+k-m)$ and $b=\frac{1}{2}(i-k)-l$ are integer numbers.
\par

On the other hand, $i-k$ even allows to replace $\eta^{i-k-2l}$ by the binomial expansion of $(1-e^2)^{\frac{1}{2}(i-k-2l)}$. Thus, replacing
\begin{equation} \label{epol}
e^{m+j}(1-e^2)^{\frac{i-k}{2}-l}
=\sum_{q=0}^{\frac{i-k}{2}-l}\binom{\frac{i-k}{2}-l}{q}(-1)^qe^{m+j+2q},
\end{equation}
into, Eq.~(\ref{Viua}), we finally get
\begin{eqnarray} 
\langle{V}_i\rangle_u &=& \sum_{l=0}^{\lfloor{i/2}\rfloor}\binom{i}{l}\binom{2i-2l}{i}\sum_{k=0}^{i-2l}\binom{i-2l}{k}\sum_{j=0}^{2l+1}\binom{2l+1}{j}\sum_{m=0}^k\binom{k}{m}\Xi_{j+k-m}^{i-2l-k} \nonumber \\ \label{Vstar}
&& \times\sum_{q=0}^{\frac{i-k}{2}-l}\binom{\frac{i-k}{2}-l}{q}(-1)^{i-l+q}e^{m+j+2q}(\hat{\Vec{e}}\cdot\hat{\Vec{r}}_\star)^k(\Vec{b}\cdot\hat{\Vec{r}}_\star)^{i-2l-k},
\end{eqnarray}
in this way making the calculation of Eq.~(\ref{H02}) complete.
\par

The generating function of the infinitesimal contact transformation leading to the averaging is computed from the usual relation
\begin{equation} \label{W2}
\mathcal{W}=\frac{1}{n}\int\left(\mathcal{V}_\star-\langle\mathcal{V}_\star\rangle_\ell\right)\mathrm{d}\ell.
\end{equation}
That is,
\begin{eqnarray}
\mathcal{W} &=& 
-\chi_{\star}L\frac{n_\star^2}{n^2}\frac{a_\star^3}{r_\star^3}\sum_{i\ge2}\frac{1}{2^i}\frac{a^{i-2}}{r_\star^{i-2}}\int\left(\frac{a}{r}V_i-\langle{V}_i\rangle_u\right)\mathrm{d}\ell  \nonumber
\\  &=& 
-\chi_{\star}L\frac{n_\star^2}{n^2}\frac{a_\star^3}{r_\star^3}\sum_{i\ge2}\frac{1}{2^i}\frac{a^{i-2}}{r_\star^{i-2}}\left(-\langle{V}_i\rangle_u\,\ell+\int V_i\mathrm{d}u\right),
\end{eqnarray}
which can be written in the form
\begin{equation}
\mathcal{W}= -\chi_{\star}L\frac{n_\star^2}{n^2}\frac{a_\star^3}{r_\star^3}\sum_{i\ge2}\frac{1}{2^i}\frac{a^{i-2}}{r_\star^{i-2}}\left[\langle{V}_i\rangle_u(u-\ell)+\int\left(V_i-\langle{V}_i\rangle_u\right)\mathrm{d}u\right],
\end{equation}
where $u-\ell=e\sin{u}$, from Kepler equation, and the integrand of the last term in the square brackets is composed only of periodic terms in the eccentric anomaly. Then, after replacing Eq.~(\ref{fourieru}) into Eq.~(\ref{Viexpanded}), we solve the indefinite integral by keeping all the trigonometric terms except those that make $i+j-m-2l-2q-2t=0$. Namely,
\begin{eqnarray} \label{periodic}
\int\left(V_i-\langle{V}_i\rangle_u\right)\mathrm{d}u &=& 
\sum_{l=0}^{\lfloor{i/2}\rfloor}\binom{i}{l}\binom{2i-2l}{i}\sum_{j=0}^{2l+1}\binom{2l+1}{j}\sum_{k=0}^{i-2l}\binom{i-2l}{k} \nonumber \\ 
&& \times\sum_{m=0}^k\binom{k}{m}e^{m+j}\eta^{i-2l-k}(\hat{\Vec{e}}\cdot\hat{\Vec{r}}_\star)^k(\Vec{b}\cdot\hat{\Vec{r}}_\star)^{i-2l-k} \nonumber\\ 
&& \times(-1)^{j+l+m}\frac{(-\ii)^{i-2l-k}}{2^{{i-2l}+{j-m}}}\sum_{ \substack{q=0 \\ q\ne{\tilde{q}}} }^{i-2l-k}\sum_{t=0}^{j+k-m}\binom{{i-2l-k}}{q} \nonumber \\ 
&& \times\binom{{j+k-m}}{t}(-1)^q\frac{\sin2(\tilde{q}-q)u-\ii\cos2(\tilde{q}-q)u}{2(\tilde{q}-q)},
\end{eqnarray}
where $\sin2(\tilde{q}-q)u-\ii\cos2(\tilde{q}-q)u=-\ii(\cos{u}+\ii\sin{u})^{2(\tilde{q}-q)}$ from which, using Eq.~(\ref{a:f2u}),
\begin{equation} 
\sin2(\tilde{q}-q)u-\ii\cos2(\tilde{q}-q)u
=-\ii\left(e+\frac{\Vec{r}\cdot\hat{\Vec{e}}}{a}+\ii\frac{\Vec{r}\cdot\Vec{b}}{a\eta}\right)^{2(\tilde{q}-q)}.
\end{equation} 
\par

Finally, the Hamiltonian in mean elements, up to higher order effects, is obtained by replacing osculating by mean elements into both the Keplerian $-\mu/(2a)$ and the third-body averaged potential, Eq.~(\ref{H02}), to yield
\begin{equation} \label{Hammean}
\mathcal{K}=-\frac{\mu}{2a}+\langle\mathcal{V}_\star\rangle_\ell,
\end{equation}
where, now, all the symbols appearing in $\mathcal{K}$ are functions of the Delaunay prime elements.
\par

\section{Hamilton equations of the averaged flow}

The flow in mean (prime) elements is obtained from the Hamilton equations. Since the mean anomaly has been removed by the perturbation approach up to the truncation order of the perturbation solution, the prime Delaunay action is a constant that decouples the reduced flow   
\begin{eqnarray} \label{dgp}
\frac{\mathrm{d}g'}{\mathrm{d}t} & = & \frac{\partial\mathcal{K}}{\partial{G}'}=\frac{\partial\langle\mathcal{V}_\star\rangle_\ell}{\partial{G}'}, \\
\frac{\mathrm{d}G'}{\mathrm{d}t} & = &-\frac{\partial\mathcal{K}}{\partial{g}'}=-\frac{\partial\langle\mathcal{V}_\star\rangle_\ell}{\partial{g}'}, \\
\frac{\mathrm{d}h'}{\mathrm{d}t} & = & \frac{\partial\mathcal{K}}{\partial{H}'}=\frac{\partial\langle\mathcal{V}_\star\rangle_\ell}{\partial{H}'},  \\  \label{dHp}
\frac{\mathrm{d}H'}{\mathrm{d}t} & = &-\frac{\partial\mathcal{K}}{\partial{h}'}=-\frac{\partial\langle\mathcal{V}_\star\rangle_\ell}{\partial{h}'},,
\end{eqnarray}
from the integration of the variation of the prime mean anomaly,
\begin{equation}
\frac{\mathrm{d}\ell'}{\mathrm{d}t}=\frac{\partial\mathcal{K}}{\partial{L}'}=n+\frac{\partial\langle\mathcal{V}_\star\rangle_\ell}{\partial{L}'},
\end{equation}
that is computed by indefinite integration after solving the reduced system (\ref{dgp})--(\ref{dHp}).
\par

In fact, in view of Eqs.~(\ref{H02}) and (\ref{Vstar}), to compute the right sides of the variation Eqs.~(\ref{dgp})--(\ref{dHp}) we only need to compute the corresponding partial derivatives of the function
\begin{equation} \label{alpha}
\alpha_{\sigma,k,p}=e^p(\hat{\Vec{e}}\cdot\hat{\Vec{r}}_\star)^k(\Vec{b}\cdot\hat{\Vec{r}}_\star)^{\sigma-k}, \qquad \sigma=i-2l, \quad p=m+j+2q,
\end{equation}
where both direction vectors $\hat{\Vec{e}}$ and $\Vec{b}$ depend on the orbital inclination $I$, the argument of the periapsis $g$ and the longitude of the node $h$, as shown in Eq.~(\ref{rotate}). 
\par

The needed partial derivatives of $\alpha_{\sigma,k,p}$ with respect to the Delaunay variables are computed using the chain rule
\begin{eqnarray}
\frac{\partial\alpha_{\sigma,k,p}}{\partial{G}} &=&
\frac{\partial\alpha_{\sigma,k,p}}{\partial{e}}\frac{\partial{e}}{\partial{G}}
+\frac{\partial\alpha_{\sigma,k,p}}{\partial(\hat{\Vec{e}}\cdot\hat{\Vec{r}}_\star)}\left(\frac{\partial\hat{\Vec{e}}}{\partial{I}}\frac{\partial{I}}{\partial{G}}\right)\cdot\hat{\Vec{r}}_\star
+\frac{\partial\alpha_{\sigma,k,p}}{\partial(\Vec{b}\cdot\hat{\Vec{r}}_\star)}\left(\frac{\partial\Vec{b}}{\partial{I}}\frac{\partial{I}}{\partial{G}}\right)\cdot\hat{\Vec{r}}_\star,
\\
\frac{\partial\alpha_{\sigma,k,p}}{\partial{H}} &=&
\frac{\partial\alpha_{\sigma,k,p}}{\partial(\hat{\Vec{e}}\cdot\hat{\Vec{r}}_\star)}\left(\frac{\partial\hat{\Vec{e}}}{\partial{I}}\frac{\partial{I}}{\partial{H}}\right)\cdot\hat{\Vec{r}}_\star
+\frac{\partial\alpha_{\sigma,k,p}}{\partial(\Vec{b}\cdot\hat{\Vec{r}}_\star)}\left(\frac{\partial\Vec{b}}{\partial{I}}\frac{\partial{I}}{\partial{H}}\right)\cdot\hat{\Vec{r}}_\star, 
\\
\frac{\partial\alpha_{\sigma,k,p}}{\partial{g}} &=&
\frac{\partial\alpha_{\sigma,k,p}}{\partial(\hat{\Vec{e}}\cdot\hat{\Vec{r}}_\star)}\frac{\partial\hat{\Vec{e}}}{\partial{g}}\cdot\hat{\Vec{r}}_\star
+\frac{\partial\alpha_{\sigma,k,p}}{\partial(\Vec{b}\cdot\hat{\Vec{r}}_\star)}\frac{\partial\Vec{b}}{\partial{g}}\cdot\hat{\Vec{r}}_\star,
\\
\frac{\partial\alpha_{\sigma,k,p}}{\partial{h}} &=&
\frac{\partial\alpha_{\sigma,k,p}}{\partial(\hat{\Vec{e}}\cdot\hat{\Vec{r}}_\star)}\frac{\partial\hat{\Vec{e}}}{\partial{h}}\cdot\hat{\Vec{r}}_\star
+\frac{\partial\alpha_{\sigma,k,p}}{\partial(\Vec{b}\cdot\hat{\Vec{r}}_\star)}\frac{\partial\Vec{b}}{\partial{h}}\cdot\hat{\Vec{r}}_\star.
\end{eqnarray}
Because $c\equiv\cos{I}=H/G$, $s\equiv\sin{I}=(1-c^2)^{1/2}$, and $e=(1-G^2/L^2)^{1/2}$, we easily get
\begin{equation} \label{partialesc}
\frac{\partial{e}}{\partial{G}}=-\frac{1}{G}\frac{\eta^2}{e}, \qquad
\frac{\partial{I}}{\partial{H}}=-\frac{1}{G}\frac{1}{s}, \qquad
\frac{\partial{I}}{\partial{G}}=\frac{1}{G}\frac{c}{s}.
\end{equation}
Besides, since the effect of a differential rotation about the axis $\Vec{n}$ (resp.~$\Vec\ell$, $\Vec{k}$) is an infinitesimal increase of the angle $g$ (resp.~$I$, $h$), we find
\begin{equation} \label{partialhate}
\frac{\partial\hat{\Vec{e}}}{\partial{g}}=\Vec{n}\times\hat{\Vec{e}}=\Vec{b}, \qquad
\frac{\partial\hat{\Vec{e}}}{\partial{h}}=\Vec{k}\times\hat{\Vec{e}}, \qquad
\frac{\partial\hat{\Vec{e}}}{\partial{I}}=\Vec\ell\times\hat{\Vec{e}}=\frac{1}{s}(\hat{\Vec{e}}\cdot\Vec{k})\Vec{n},
\end{equation}
and
\begin{equation} \label{partialvecb}
\frac{\partial\Vec{b}}{\partial{g}}=\Vec{n}\times\Vec{b}=-\hat{\Vec{e}}, \qquad
\frac{\partial\Vec{b}}{\partial{h}}=\Vec{k}\times\Vec{b}, \qquad
\frac{\partial\Vec{b}}{\partial{I}}=\Vec\ell\times\Vec{b}=\frac{1}{s}(\Vec{b}\cdot\Vec{k})\Vec{n}.
\end{equation}
\par

Thus, straightforward computations yield
\begin{eqnarray} \label{pdalG}
\frac{\partial\alpha_{\sigma,k,p}}{\partial{G}} &=& 
-\frac{1}{G}\frac{\eta^2}{e}\left[pe^{p-1}\right](\hat{\Vec{e}}\cdot\hat{\Vec{r}}_\star)^{k}(\Vec{b}\cdot\hat{\Vec{r}}_\star)^{\sigma-k} \nonumber\\ 
&&
+\frac{1}{G}\frac{c}{s}e^{p}\left[k(\hat{\Vec{e}}\cdot\hat{\Vec{r}}_\star)^{k-1}\right](\Vec{b}\cdot\hat{\Vec{r}}_\star)^{\sigma-k}(\Vec{n}\cdot\hat{\Vec{r}}_\star)\frac{\hat{\Vec{e}}\cdot\Vec{k}}{s} \nonumber\\ 
&&
+\frac{1}{G}\frac{c}{s}e^{p}(\hat{\Vec{e}}\cdot\hat{\Vec{r}}_\star)^{k}\left[(\sigma-k)(\Vec{b}\cdot\hat{\Vec{r}}_\star)^{\sigma-k-1}\right](\Vec{n}\cdot\hat{\Vec{r}}_\star)\frac{\Vec{b}\cdot\Vec{k}}{s},
\\[1.ex] \label{pdalH}
\frac{\partial\alpha_{\sigma,k,p}}{\partial{H}} &=&
-\frac{1}{G}\frac{1}{s}e^{p}\left[k(\hat{\Vec{e}}\cdot\hat{\Vec{r}}_\star)^{k-1}\right](\Vec{b}\cdot\hat{\Vec{r}}_\star)^{\sigma-k}(\Vec{n}\cdot\hat{\Vec{r}}_\star)\frac{\hat{\Vec{e}}\cdot\Vec{k}}{s} \nonumber\\ 
&& -\frac{1}{G}\frac{1}{s}e^{p}(\hat{\Vec{e}}\cdot\hat{\Vec{r}}_\star)^{k}\left[(\sigma-k)(\Vec{b}\cdot\hat{\Vec{r}}_\star)^{\sigma-k-1}\right](\Vec{n}\cdot\hat{\Vec{r}}_\star)\frac{\Vec{b}\cdot\Vec{k}}{s},
\\[1.ex] \label{pdalw}
\frac{\partial\alpha_{\sigma,k,p}}{\partial{g}} &=&
e^p\left\{\left[k(\hat{\Vec{e}}\cdot\hat{\Vec{r}}_\star)^{k-1}\right](\Vec{b}\cdot\hat{\Vec{r}}_\star)^{\sigma-k+1} \right. \nonumber\\ 
&& \left. -(\hat{\Vec{e}}\cdot\hat{\Vec{r}}_\star)^{k+1}\left[(\sigma-k)(\Vec{b}\cdot\hat{\Vec{r}}_\star)^{\sigma-k-1}\right]\right\},
\\[1.ex] \label{pdalO}
\frac{\partial\alpha_{\sigma,k,p}}{\partial{h}} &=&
e^p\left\{\left[k(\hat{\Vec{e}}\cdot\hat{\Vec{r}}_\star)^{k-1}\right](\Vec{b}\cdot\hat{\Vec{r}}_\star)^{\sigma-k}(\hat{\Vec{e}}\times\hat{\Vec{r}}_\star) \right. \nonumber\\ 
&& \left. +(\hat{\Vec{e}}\cdot\hat{\Vec{r}}_\star)^k\left[(\sigma-k)(\Vec{b}\cdot\hat{\Vec{r}}_\star)^{\sigma-k-1}\right](\Vec{b}\times\hat{\Vec{r}}_\star) \right\}\cdot\Vec{k},
\end{eqnarray}
where the unneeded square brackets in these equations are used to emphasize that corresponding enclosed terms never introduce denominators.

\section{Long-term flow in mean vectorial elements}

Note that, as expected from the singularities of Delaunay variables, Eqs.~(\ref{pdalG}) and (\ref{pdalH}) are singular in the case of circular and equatorial orbits. However, these singularities are of virtual nature \citep{Henrard1974} and may be removed when the equations of the flow are represented in other set of canonical or non-canonical variables. In particular, the variations experienced by the angular momentum vector and the eccentricity vector under third-body perturbations are free from singularities, and provide a general, compact, and elegant way of presenting the mean elements' variations. 
\par

Indeed, scaling the mean angular momentum vector by $L'$ the variation of the mean vectorial elements $\Vec{h}=\Vec{G}/L'=\eta\Vec{n}$ and $\Vec{e}=e\hat{\Vec{e}}$ take the neat, symmetric form \citep{Milankovitch1941,AllanWard1963,AllanCook1964,RosengrenScheeres2014}
\begin{eqnarray} \label{dhdt}
\frac{\mathrm{d}\Vec{h}}{\mathrm{d}t} &=& \Vec{h}\times\frac{\partial\mathcal{R}}{\partial\Vec{h}} 
+ \Vec{e}\times\frac{\partial\mathcal{R}}{\partial\Vec{e}}, \\ \label{dedt}
\frac{\mathrm{d}\Vec{e}}{\mathrm{d}t} &=& \Vec{e}\times\frac{\partial\mathcal{R}}{\partial\Vec{h}} 
+ \Vec{h}\times\frac{\partial\mathcal{R}}{\partial\Vec{e}},
\end{eqnarray}
in which
\begin{equation}
\mathcal{R}=-\frac{1}{L'}\langle\mathcal{V}_\star\rangle_\ell,
\end{equation}
and $\langle\mathcal{V}_\star\rangle_\ell$ is given in Eq.~(\ref{H02}). Furthermore, derivation of Eqs.~(\ref{dhdt})--(\ref{dedt}) is trivial after minor rearrangement of Eq.~(\ref{Vstar}) to replace the directions $\hat{\Vec{e}}$ and $\Vec{n}$ by the non-dimensional magnitudes $\Vec{e}$ and $\Vec{h}$, respectively. 
\par

Thus, instead of replacing Eq.~(\ref{epol}) in Eq.~(\ref{Vstar}), the latter is written in the more convenient  form
\begin{eqnarray} 
\langle{V}_i\rangle_u &=& \sum_{l=0}^{\lfloor{i/2}\rfloor}\binom{i}{l}\binom{2i-2l}{i}\sum_{j=0}^{2l+1}\binom{2l+1}{j}\sum_{k=0}^{i-2l}\binom{i-2l}{k}\sum_{m=0}^k\binom{k}{m} \nonumber\\ 
&& \times \Xi_{j+k-m}^{i-2l-k}(-1)^{j+l+m}e^{j+m-k}(e\hat{\Vec{e}}\cdot\hat{\Vec{r}}_\star)^k\left[\eta^2(\Vec{b}\cdot\hat{\Vec{r}}_\star)^2\right]^{\frac{i-k}{2}-l}.
\end{eqnarray}
If now we make use of the identity
\begin{equation} \label{projectionsrstar}
(\hat{\Vec{e}}\cdot\hat{\Vec{r}}_\star)^2+(\Vec{b}\cdot\hat{\Vec{r}}_\star)^2+(\Vec{n}\cdot\hat{\Vec{r}}_\star)^2=1,
\end{equation}
that gives the square of the modulus of the third-body direction $\|\hat{\Vec{r}}_\star\|=1$ when computed in the apsidal frame, the dependence of the averaged potential on the unit vector $\Vec{b}$ in the binormal direction is replaced by a corresponding dependence on the normal direction $\Vec{n}$. That is
\begin{eqnarray} 
\langle{V}_i\rangle_u &=& \sum_{l=0}^{\lfloor{i/2}\rfloor}\binom{i}{l}\binom{2i-2l}{i}\sum_{j=0}^{2l+1}\binom{2l+1}{j}\sum_{k=0}^{i-2l}\binom{i-2l}{k}\sum_{m=0}^k\binom{k}{m} \nonumber \\
&& \times \Xi_{j+k-m}^{i-2l-k}(-1)^{j+l+m}(\Vec{e}\cdot\Vec{e})^{\frac{j+m-k}{2}}(\Vec{e}\cdot\hat{\Vec{r}}_\star)^k\left[X-(\hat{\Vec{e}}\cdot\hat{\Vec{r}}_\star)^2\right]^{\frac{i-k}{2}-l},
\end{eqnarray}
with the scalar function
\begin{equation}
X=X(\Vec{e},\Vec{h},\hat{\Vec{r}}_\star)\equiv1-e^2+\xi^2-\zeta^2,
\end{equation}
where we introduced the abbreviations $\xi=\Vec{e}\cdot\hat{\Vec{r}}_\star$, $\zeta=\Vec{h}\cdot\hat{\Vec{r}}_\star$.
\par

Now, applying once more the binomial expansion we find
\begin{eqnarray} \nonumber
\langle{V}_i\rangle_u &=& \sum_{l=0}^{\lfloor{i/2}\rfloor}\binom{i}{l}\binom{2i-2l}{i}\sum_{j=0}^{2l+1}\binom{2l+1}{j}\sum_{k=0}^{i-2l}\binom{i-2l}{k}\sum_{m=0}^k\binom{k}{m}\sum_{s=0}^{\frac{i-k}{2}-l}\binom{\frac{i-k}{2}-l}{s} \\ \label{Viaveh}
&& \times \Xi_{j+k-m}^{i-2l-k}(-1)^{j+l+m+s}(\Vec{e}\cdot\Vec{e})^{\frac{j+m-k}{2}-s}(\Vec{e}\cdot\hat{\Vec{r}}_\star)^{2s+k}X^{\frac{i-k}{2}-l-s},
\end{eqnarray}
that only depends on $\Vec{e}$, and $\Vec{h}$, as desired, as well as on the disturbing body direction $\hat{\Vec{r}}_\star$.
\par 

Explicit expressions for the first terms of the Legendre polynomials expansion yield the compact expressions
\begin{eqnarray}
\langle{V}_2\rangle_u &=& 1-3\left(2e^2-5\xi^2+\zeta^2\right),
\\
\langle{V}_3\rangle_u &=& \mbox{$-\frac{5}{2}$}\xi\left[3\left(1-8^2\right)-15\zeta^2+35\xi^2\right], 
\\
\langle{V}_4\rangle_u &=& \mbox{$\frac{3}{4}$}\left\{3 - 20 e^2 + 80 e^4 
+ 10\left[7\left(1-10e^2\right)\xi^2-\left(3-10e^2\right)\zeta^2\right] \right. \nonumber \\
&& \left. +35\left(21\xi^4-14\zeta^2\xi^2+\zeta^4\right) \right\},
\\
\langle{V}_5\rangle_u &=& \mbox{$-\frac{3}{4}$}\xi\left\{35\left(1-8e^2+40e^4\right)
+490\left[\left(1-12e^2\right)\xi^2 -\left(1-4e^2\right)\zeta^2\right] \right. \nonumber \\
&& \left. +147\left(33\xi^4-30\zeta^2\xi^2+5\zeta^4\right)\right\},
\\
\langle{V}_6\rangle_u &=& \mbox{$\frac{1}{4}$}\left\{5\left(5-42e^2+168e^4-560e^6\right) \right. \nonumber \\
&& +105\left[3\left(3-28e^2+168e^4\right)\xi^2 -\left(5-28e^2+56e^4\right)\zeta^2\right] \nonumber \\
&& +315\left[\left(5-14e^2\right)\zeta^4-18\left(3-14e^2\right)\zeta^2\xi^2+33\left(1-14e^2\right)\xi^4\right] \nonumber \\
&& \left. +231\left(429\xi^6-495\zeta^2\xi^4+135\zeta^4\xi^2-5\zeta^6\right) \right\},
\\
\langle{V}_7\rangle_u &=& \mbox{$-\frac{9}{16}$}\xi\left\{35\left(5-48e^2+224e^4-896e^6\right) \right. \nonumber \\
&& +105\left[11\left(3-32e^2+224e^4\right)\xi^2 -3\left(15-96e^2+224e^4\right) \zeta^2\right] \nonumber \\
&& +231\left[143\left(1-16e^2\right)\xi^4 -110\left(3-16e^2\right)\zeta^2\xi^2 +15\left(5-16e^2\right)\zeta^4\right] \nonumber \\
&& \left.+429\left(715\xi^6-1001\zeta^2\xi^4+385\zeta^4\xi^2-35\zeta^6\right) \right\}, \\
\langle{V}_8\rangle_u &=& \mbox{$\frac{5}{64}$}\left\{ 7\left(35-360e^2+1728e^4-5376e^6+16128e^8\right)
+252\left[11\left(5  \right. \right. \right. \nonumber \\
&& \left. \left. -54e^2+288e^4-1344e^6\right)\xi^2 -\left(35-270e^2+864e^4-1344e^6\right)\zeta^2\right] \nonumber \\
&& +1386 \left[143 \left(1 - 12 e^2 + 96 e^4\right) \xi ^4
-66 \left(5 - 36 e^2 + 96 e^4\right) \zeta ^2 \xi ^2 \right. \nonumber \\
&& \left. +\left(35 - 180 e^2 + 288 e^4\right) \zeta ^4
\right]
+12012\left[ 143\left(1-18e^2\right)\xi^6 \right. \nonumber 
\\ && \left.
+33 \left(5-18e^2\right)\xi^2\zeta^4-429\left(1-6e^2\right)\xi^4\zeta^2-\left(7-18e^2\right)\zeta^6
\right] \nonumber \\
&& \left. +6435 \left(2431\xi^8-4004\zeta^2\xi^6+2002\zeta^4\xi^4-308\zeta^6\xi^2+7\zeta^8\right)
\right\}.
\end{eqnarray}
The terms $\langle{V}_2\rangle_u$ and $\langle{V}_3\rangle_u$ have been repeatedly reported in the literature \citep{Musen1961,RosengrenScheeres2013} whereas the remaining terms have been checked with corresponding expressions in Delaunay elements customarily used in orbit propagators based on semi-analytical integration \citep{LaraSanJuanLopezCefola2012,LaraSanJuanHautesserresCNES2016,LaraSanJuanHautesserres2018}. Note that the constant term in $\langle{V}_2\rangle_u$ is commonly neglected because it has no effects in the long term motion.
\par

Finally, on account of
\begin{eqnarray}
\frac{\partial(\Vec{e}\cdot\Vec{e})}{\partial\Vec{h}} &=& \Vec{0}, \\
\frac{\partial(\Vec{e}\cdot\Vec{e})}{\partial\Vec{e}} &=& 2\Vec{e}, \\
\frac{\partial(\Vec{e}\cdot\hat{\Vec{r}}_\star)}{\partial\Vec{h}} &=& \Vec{0}, \\
\frac{\partial(\Vec{e}\cdot\hat{\Vec{r}}_\star)}{\partial\Vec{e}} &=& \hat{\Vec{r}}_\star, \\
\frac{\partial{X}}{\partial\Vec{h}} &=& -2(\Vec{h}\cdot\hat{\Vec{r}}_\star)\hat{\Vec{r}}_\star, \\
\frac{\partial{X}}{\partial\Vec{e}} &=& -2\Vec{e}+2(\Vec{e}\cdot\hat{\Vec{r}}_\star)\hat{\Vec{r}}_\star,
\end{eqnarray}
derivation of Eqs.~(\ref{dhdt})--(\ref{dedt}) is straightforward from Eq.~(\ref{Viaveh}).
\par

Therefore, using Eq.~(\ref{H02}) and taking into account that $(\mu/2a)/L=\frac{1}{2}n$, Eqs.~(\ref{dhdt})--(\ref{dedt}) are written in the form
\begin{eqnarray} \label{dhdtsum}
\frac{\mathrm{d}\Vec{h}}{\mathrm{d}t} &=& n\epsilon\sum_{i\ge2}\frac{1}{2^i}\frac{a^{i-2}}{r_\star^{i-2}}\left[\gamma_i(\Vec{h}\times\hat{\Vec{r}}_\star)+\rho_i(\Vec{e}\times\hat{\Vec{r}}_\star)\right],%\dot{\Vec{h}}_i, 
\\ \label{dedtsum}
\frac{\mathrm{d}\Vec{e}}{\mathrm{d}t} &=& n\epsilon\sum_{i\ge2}\frac{1}{2^i}\frac{a^{i-2}}{r_\star^{i-2}}\left[\gamma_i(\Vec{e}\times\hat{\Vec{r}}_\star)+\rho_i(\Vec{h}\times\hat{\Vec{r}}_\star)+4\rho_{i-1}(\Vec{h}\times\Vec{e})\right],%\dot{\Vec{e}}_i,
\end{eqnarray}
where we abbreviate $\epsilon=\chi_{\star}(n_\star/n)^2(a_\star/r_\star)^3$, the polynomials $\rho_i$, are computed from the recursion
\begin{eqnarray} \nonumber
\rho_i &=& \sum_{l=0}^{\lfloor{i/2}\rfloor}\binom{i}{l}\binom{2i-2l}{i}\sum_{j=0}^{2l+1}\binom{2l+1}{j}\sum_{k=0}^{i-2l}\binom{i-2l}{k}\sum_{m=0}^k\binom{k}{m}\sum_{s=0}^{\frac{i-k}{2}-l}\binom{\frac{i-k}{2}-l}{s} \\ \nonumber
&& \times\Xi_{j+k-m}^{i-2l-k}(-1)^{j+l+m+s}e^{j+m-k-2s}\xi^{2s+k-1}X^{\frac{i-k}{2}-l-s-1} \\ \label{Pi}
&& \times\left[(k+2s)X+(i-k-2l-2s)\xi^2\right],
\end{eqnarray}
and those $\gamma_i$ from
\begin{eqnarray} 
\gamma_i &=& \sum_{l=0}^{\lfloor{i/2}\rfloor}\binom{i}{l}\binom{2i-2l}{i}\sum_{j=0}^{2l+1}\binom{2l+1}{j}\sum_{k=0}^{i-2l}\binom{i-2l}{k}\sum_{m=0}^k\binom{k}{m}\sum_{s=0}^{\frac{i-k}{2}-l}\binom{\frac{i-k}{2}-l}{s} \nonumber\\ 
&& \times\Xi_{j+k-m}^{i-2l-k}(-1)^{j+l+m+s+1}(i-k-l-s)e^{j+m-k-2s}\xi^{2s+k}\zeta{X}^{\frac{i-k}{2}-l-s-1}. \label{Qi}
\end{eqnarray}
The complexity of the polynomials given by Eqs.~(\ref{Pi}) and (\ref{Qi}) is only apparent, and corresponding expressions take a compact form, which is illustrated in Table \ref{t:iep} for the lower degrees.

\begin{table*}[htbp]
\caption{Some polynomials $\rho_i$, $\gamma_i$, given by Eqs.~(\protect\ref{Pi})--(\protect\ref{Qi}).}
\centering
\begin{tabular}{@{}lcl@{}}
\hline
$\rho_1$ & $=$ & $-3$, \\[0.5ex]
$\rho_2$ & $=$ & $30\xi$, \\[0.5ex]
$\gamma_2$ & $=$ & $-6\zeta$, \\[0.5ex]
$\rho_3$ & $=$ & $-\mbox{$\frac{15}{2}$}\left[1-8e^2+5\left(7\xi^2-\zeta^2\right)\right]$, \\[0.5ex]
$\gamma_3$ & $=$ & $75\xi\zeta$, \\[0.5ex]
$\rho_4$ & $=$ & $105\left[1-10e^2+7\left(3\xi^2-\zeta^2\right)\right]\xi$, \\[0.5ex]
$\gamma_4$ & $=$ & $-15\left[3-10e^2+7\left(7\xi^2-\zeta^2\right)\right]\zeta$, \\[0.5ex]
$\rho_5$ & $=$ & $-\mbox{$\frac{105}{4}$}\left\{1-8e^2+40e^4+14\left[3\left(1-12e^2\right)\xi^2-\left(1-4e^2\right)\zeta^2\right] \right.$ \\
      & & $\left. +21\left(33\xi^4-18\xi^2\zeta^2+\zeta^4\right)\right\}$, \\[0.5ex]
$\gamma_5$ & $=$ & $735\left(1-4e^2+9\xi^2-3\zeta^2\right)\xi\zeta$, \\[0.5ex]
$\rho_6$ & $=$ & $\mbox{$\frac{63}{2}$}\left\{5\left(3-28e^2+168e^4\right)+30\left[11\left(1-14e^2\right)\xi^2-3\left(3-14 e^2\right)\zeta^2\right] \right.$ \\
      & & $\left. +33\left(143\xi^4-110\xi^2\zeta^2+15\zeta^4\right)\right\}\xi$, \\[0.5ex]
$\gamma_6$ & $=$ & $-\mbox{$\frac{105}{2}$}\left\{5-28e^2+56e^4+6\left[9\left(3-14e^2\right)\xi^2-\left(5-14e^2\right)\zeta^2\right] \right.$ \\
      & & $\left. +33\left(33\xi^4-18\xi^2\zeta^2+\zeta^4\right)\right\}\zeta$, \\[0.5ex]
$\rho_7$ & $=$ & $-\mbox{$\frac{315}{16}$}\left\{5-48e^2+224e^4-896e^6 \right.$ \\
      & & $+9\left[11\left(3-32e^2+224e^4\right)\xi^2-\left(15-96e^2+224e^4\right)\zeta^2\right]$ \\
      & & $+33\left[143\left(1-16e^2\right)\xi^4-66\left(3-16e^2\right)\xi^2\zeta^2+3\left(5-16e^2\right)\zeta^4\right]$ \\
      & & $ \left. +429\left(143\xi^6-143\xi^4\zeta^2+33\xi^2\zeta^4-\zeta^6\right) \right\}$, \\[0.5ex]
$\gamma_7$ & $=$ & $\mbox{$\frac{189}{8}$}\left\{ 15\left(15-96 e^2+224 e^4\right) \right.$ \\
& & $\left. +110 \left[11 \left(3-16 e^2\right)\xi^2-3 \left(5-16 e^2\right) \zeta^2\right] \right.$ \\
      & & $ \left. +143\left(143 \xi^4-110 \xi^2\zeta^2+15\zeta^4\right) \right\}\xi\zeta$, \\[0.5ex]
$\rho_8$ & $=$ & $\mbox{$\frac{495}{8}$} \left\{7\left(5 - 54 e^2 + 288 e^4 - 1344 e^6\right) \right.$ \\
& & $+ 77 \left[13 \left(1 - 12 e^2 + 96 e^4\right) \xi^2 - 
      3 \left(5 - 36 e^2 + 96 e^4\right) \zeta^2\right]$ \\ 
& & $+ 1001 \left[13 \left(1 - 18 e^2\right) \xi^4 - 
      26 \left(1 - 6 e^2\right) \xi^2 \zeta^2 + \left(5 - 18 e^2\right) \zeta^4\right]$ \\
& & $\left. + 715 \left(221 \xi^6 - 273 \xi^4 \zeta^2 + 91 \xi^2 \zeta^4 - 
      7 \zeta^6\right)\right\} \xi$, \\[0.5ex]
$\gamma_8$ & $=$ & $-\mbox{$\frac{315}{8}$}\left\{35-270e^2+864e^4-1344e^6 \right. $ \\
& & $+11\left[33\left(5-36e^2+96e^4\right)\xi^2-\left(35-180e^2+288e^4\right)\zeta^2\right] $ \\
& & $+143\left[143\left(1-6e^2\right)\xi^4-22\left(5-18e^2\right)\xi^2\zeta^2+\left(7-18e^2\right) \zeta^4\right]$ \\
& & $\left. +715\left(143\xi^6-143\xi^4\zeta^2+33\xi^2\zeta^4-\zeta^6\right)\right\}\zeta.$ \\
\hline
\end{tabular}
\label{t:iep}
\end{table*}

\section{Performance evaluation: The case of high-Earth orbits}

The simplicity of the vectorial approach in approximating the long-term dynamics of a system under third-body perturbations when compared with classical expansions based in trigonometric terms is evident from the visual comparison of Table \ref{t:iep} with equivalent results based on classical, trigonometric expansions. Thus, for instance, while the entire Tables 4-8 of \cite{LaraSanJuanHautesserres2018} are needed in the evaluation of the third-body disturbing acceleration up to the degree 6 of the Legendre polynomial expansion ---which, besides, require the additional evaluation of trigonometric functions--- it only requires the upper half of Table \ref{t:iep} when the vectorial approach is used. On the other hand, the numerical integration of the mean elements is expected to progress with similar step sizes when using either vectorial or other classical mean elements. Still, the former requires the integration of a higher dimension, redundant differential system, a fact that might counterbalance the presumed advantages derived from the simplicity of the formulation and the only use of arithmetic operations, as opposite to the evaluation of trigonometric functions. 
\par

The improvements obtained with the vectorial formulation when higher degrees are needed in the expansion of the third-body disturbing function are illustrated for high-Earth orbits. In particular, we present an example of a SIMBOL-X-type orbit, whose semi-analytical propagation requires the use of at least a $P_2$--$P_6$ truncation of the lunar disturbing function in order to capture the main frequencies of the long-term motion \citep{LaraSanJuanHautesserres2018,Amato2019}. 
\par

For our efficiency proofs, we take the initial conditions
\[
\begin{array}{ccl}
a & = & 106\,247.136\,\mathrm{km} \\
e & = & 0.75173  \\
I & = & 5.2789^{\circ}  \\
\Omega & = & 49.351^{\circ} \\
\omega & = & 180.008^{\circ} \\
M & = & 0  \\
\end{array}
\]
where $a$, $e$, $I$, $\Omega$, $\omega$, and $M$, stand for classical Keplerian elements, and remove all perturbations except for the lunisolar ones. The initial epoch needed for the computation of lunisolar ephemeris is fixed to July 1, 2014, at $20.7208333$ h UTC. Then, we propagate these initial conditions for intervals of 100 years with both the vectorial approach of this paper and with the classical approach used by \cite{LaraSanJuanHautesserres2018}. The solar effect is truncated to the $P_2$ Legendre polynomial whereas lunar perturbations take from $P_2$ alone to $P_2$--$P_8$ into account in successive increments of one term. The numerical integration of the averaged flow is carried out with the reputed DOPRI 853 integrator, which implements an explicit Runge-Kutta method of order 8(5,3) due to  \cite{DormandPrince1980}, with step size control and dense output \citep{HairerNorsettWanner2008}.
\par

\begin{figure}[htbp]
\centering
\includegraphics[scale=0.9]{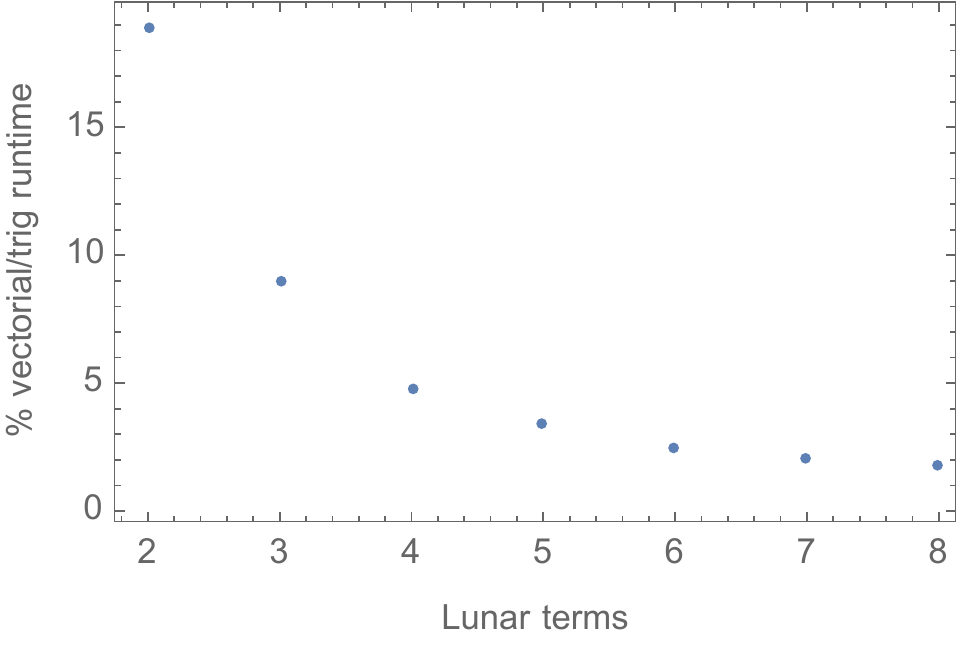} 
\caption{Runtime of the vectorial approach relative to the classical trigonometric implementation, in percentage, for increasing complexity of the lunar perturbation.}
\label{f:v2tpercentage}
\end{figure}

In order to estimate the gains of the vectorial formulation with respect to the trigonometric expansions, we first make the propagations with constant step size and without updating the lunisolar ephemerides. The results are depicted in Fig.~\ref{f:v2tpercentage} in terms of the runtime percentage of the vectorial approach relative to the classical propagation. Since the figures presented are relative quantities, analogous results are expected using different computational environments and, therefore, there is no need of providing additional information on the hardware and software used in the tests. It is worth, however, mentioning that we declined to optimize the evaluation of the disturbing acceleration and, except for trivial arrangements, we completely left the optimization task to the compiler in both cases ---the vectorial formulation of this paper and the classical approach of \cite{LaraSanJuanHautesserres2018}.
\par

As shown in Fig.~\ref{f:v2tpercentage}, when the lunar perturbation is limited to the quadrupolar term the vectorial formulation only needs about one fifth of the time needed by the classical formulation in trigonometric functions to complete the 100 years propagation. The gains of the vectorial formulation notably increase with the fidelity in modeling lunar perturbations. Indeed, the vectorial approach performs in less than one tenth of the classical approach of \cite{LaraSanJuanHautesserres2018} when the lunar octupolar term is also taken into account, and achieves an impressive $1.8\%$ of the time needed by the classical formulation when the terms $P_2$--$P_8$ are taken into account. Recall that for this particular orbit the long-term dynamics is correctly modeled with a $P_2$--$P_6$ truncation of the lunar disturbing function, cf.~\cite{LaraSanJuanHautesserres2018}, a case in which the runtime obtained with the vectorial formulation is less than the $2.5\%$ of the time needed by the trigonometric expansions.
\par

\begin{figure}[htbp]
\centering
\includegraphics[scale=0.9]{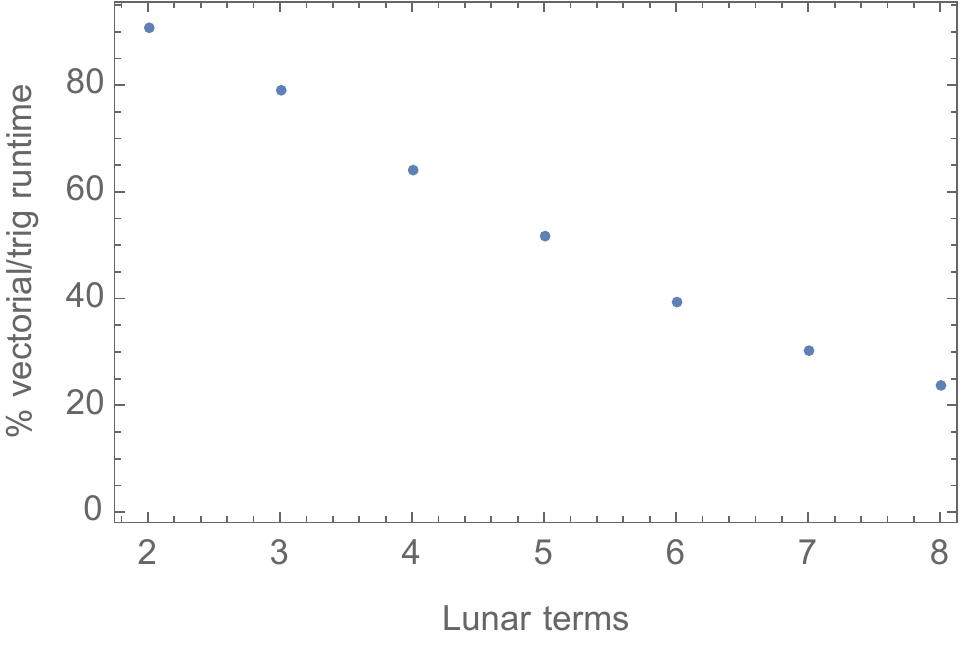} 
\caption{Same as Fig.~\protect\ref{f:v2tpercentage} when the time spent in the evaluation of lunisolar ephemeris is included.}
\label{f:v2tpercentageFull}
\end{figure}

On the other hand, the evaluation of the third-body acceleration, as given by Eqs.~(\ref{dhdtsum})--(\ref{dedtsum}) or equivalent expressions in \cite{LaraSanJuanHautesserres2018}, is only a part of the computational burden of the propagation. Indeed, the evaluation of the disturbing acceleration requires the previous knowledge of the third body's ephemeris at each step of the numerical integration of the flow in mean elements. While this task may be computationally undemanding for simple restricted three-body models in which the orbit of the disturber may be taken as purely Keplerian \citep{LaraSanJuanLopezCefola2012}, it is not at all the case for Earth orbiting satellites, in which lunar and solar ephemeris must either be read from a data file or evaluated from analytical expressions. In this last case the added computational effort is not negligible at all, and, therefore, it naturally emerges the question of whether or not the efforts in improving evaluation of the differential equations of the flow might be vacuous. To check that we repeat the computations in the actual case in which lunisolar ephemeris are computed at each integration step of the numerical integration procedure. Corresponding results are shown in Fig.~\ref{f:v2tpercentageFull}. Now, runtime improvements are quite small for the quadrupolar case and just moderate for the octupolar one. Still, the vectorial approach halves runtime when moderate degrees are needed in the expansion of the lunar disturbing function, and it only needs about 20\% of the runtime required by the classical approach when the truncation is extended to the $P_2$--$P_8$ case.

Finally, we recall that an added bonus on the side of the vectorial formulation is that its redundancy allows to examine the accuracy of the numerical integration by the evaluation of the geometrical relation $\Vec{h}\cdot\Vec{e}=0$ and $\Vec{e}\cdot\Vec{e}+\Vec{h}\cdot\Vec{h}=1$, cf.~\citep{Herrick1948}. Using these tests, we found that the errors of the 100 year numerical propagation of the flow in mean vectorial elements is constrained to the order of $10^{-14}$ when the tolerance of the numerical integrator is set to $10^{-14}$.

\section{Conclusions}

We have provided a detailed description of the Hamiltonian reduction process that removes short-period effects from the third-body disturbing function, keeping the slowly evolving parameters in vectorial form. Our approach takes advantage of the apsidal frame formulation to provide general expressions of the mean elements potential in vectorial form up to an arbitrary degree. The variation of parameter equations of the flow in mean elements admit a general, non-singular, compact formulation when the elements are the angular momentum vector and the eccentricity vector. Explicit vectorial expressions for the lower degree truncations of our general approach have been checked to agree with alternative derivations in the literature, and can now be extended to an arbitrary degree. Besides, the compact form of the vectorial expressions permits to avoid the long listings required when using angular variables, and printed expressions up to moderate orders of the variation equations can be arranged in barely one side of a letter. Runs for particular examples confirmed the expected, notable improvements of the performance in the propagation of the mean elements' flow when using the vectorial approach.

\begin{acknowledgements}
The work of ML and EF has been funded by Khalifa University of Science and Technology's internal grants FSU-2018-07 and CIRA-2018-85. 
ML also thanks partial support by the Spanish State Research Agency and the European Regional Development Fund under Projects ESP2016-76585-R and ESP2017-87271-P (AEI/ ERDF, EU).
\end{acknowledgements}

\bibliographystyle{aa}
\bibliography{references}   % Use bibref.bib to resolve the labels.

\begin{thebibliography}{55}
\expandafter\ifx\csname natexlab\endcsname\relax\def\natexlab#1{#1}\fi

\bibitem[{{Allan}(1962)}]{Allan1962}
{Allan}, R.~R. 1962, The Quarterly Journal of Mechanics and Applied
  Mathematics, 15, 283

\bibitem[{{Allan} \& {Cook}(1964)}]{AllanCook1964}
{Allan}, R.~R. \& {Cook}, G.~E. 1964, Proceedings of the Royal Society of
  London Series A, 280, 97

\bibitem[{{Allan} \& {Ward}(1963)}]{AllanWard1963}
{Allan}, R.~R. \& {Ward}, G.~N. 1963, Proceedings of the Cambridge
  Philosophical Society, 59, 669

\bibitem[{{Amato} {et~al.}(2019){Amato}, {Bombardelli}, {Ba{\'u}}, {Morand}, \&
  {Rosengren}}]{Amato2019}
{Amato}, D., {Bombardelli}, C., {Ba{\'u}}, G., {Morand}, V., \& {Rosengren},
  A.~J. 2019, Celestial Mechanics and Dynamical Astronomy, 131, 21

\bibitem[{{Andrade-Ines} {et~al.}(2016){Andrade-Ines}, {Beaug\'e},
  {Michtchenko}, \& {Robutel}}]{Andrade-Inesetal2016}
{Andrade-Ines}, E., {Beaug\'e}, C., {Michtchenko}, T., \& {Robutel}, P. 2016,
  Celestial Mechanics and Dynamical Astronomy, 124, 405

\bibitem[{{Andrade-Ines} \& {Robutel}(2018)}]{Andrade-InesRobutel2018}
{Andrade-Ines}, E. \& {Robutel}, P. 2018, Celestial Mechanics and Dynamical
  Astronomy, 130, 6

\bibitem[{{Beaug\'e} \& {Michtchenko}(2003)}]{BeaugeMichtchenko2003}
{Beaug\'e}, C. \& {Michtchenko}, T.~A. 2003, Monthly Notices of the Royal
  Astronomical Society, 341, 760

\bibitem[{Blanes {et~al.}(2013)Blanes, Casas, Farr\'es, Laskar, Makazaga, \&
  Murua}]{Blanesetal2013}
Blanes, S., Casas, F., Farr\'es, A., {et~al.} 2013, Applied Numerical
  Mathematics, 68, 58

\bibitem[{{Boccaletti} \& {Pucacco}(2002)}]{BoccalettiPucacco1998v2}
{Boccaletti}, D. \& {Pucacco}, G. 2002, {Theory of orbits. Volume 2:
  Perturbative and geometrical methods}, 1st edn., Astronomy and Astrophysics
  Library (Berlin Heidelberg New York: Springer-Verlag)

\bibitem[{{Breiter} \& {Ratajczak}(2005)}]{BreiterRatajczak2005}
{Breiter}, S. \& {Ratajczak}, R. 2005, Monthly Notices of the Royal
  Astronomical Society, 364, 1222

\bibitem[{{Celletti} {et~al.}(2017){Celletti}, {Gale\c{s}}, {Pucacco}, \&
  {Rosengren}}]{Cellettietal2017}
{Celletti}, A., {Gale\c{s}}, C., {Pucacco}, G., \& {Rosengren}, A.~J. 2017,
  Celestial Mechanics and Dynamical Astronomy, 127, 259

\bibitem[{{Correia} {et~al.}(2011){Correia}, {Laskar}, {Farago}, \&
  {Bou{\'e}}}]{Correiaetal2011}
{Correia}, A. C.~M., {Laskar}, J., {Farago}, F., \& {Bou{\'e}}, G. 2011,
  Celestial Mechanics and Dynamical Astronomy, 111, 105

\bibitem[{{Deprit}(1969)}]{Deprit1969}
{Deprit}, A. 1969, Celestial Mechanics, 1, 12

\bibitem[{{Deprit}(1975)}]{Deprit1975}
{Deprit}, A. 1975, Journal of Research of the National Bureau of Standards, 79,
  1

\bibitem[{{Deprit}(1982)}]{Deprit1982}
{Deprit}, A. 1982, Celestial Mechanics, 26, 9

\bibitem[{{Deprit}(1983)}]{Deprit1983}
{Deprit}, A. 1983, Celestial Mechanics, 29, 229

\bibitem[{{Deprit}(1984)}]{Deprit1984}
{Deprit}, A. 1984, in The Big-Bang and Georges Lema\^{i}tre, ed. A.~{Berger}
  (Dordrecht: Springer), 151--180

\bibitem[{Dormand \& Prince(1980)}]{DormandPrince1980}
Dormand, J.~R. \& Prince, P.~J. 1980, Journal of Computational and Applied
  Mathematics, 6, 19

\bibitem[{{Giacaglia}(1974)}]{Giacaglia1974}
{Giacaglia}, G.~E.~O. 1974, Celestial Mechanics, 9, 239

\bibitem[{Hairer {et~al.}(2008)Hairer, N{\o}rset, \&
  Wanner}]{HairerNorsettWanner2008}
Hairer, E., N{\o}rset, S.~P., \& Wanner, G. 2008, {Solving Ordinary
  Differential Equations I. Non-stiff Problems}, 2nd edn. (Berlin -- Heidelberg
  -- New York: Springer-Verlag)

\bibitem[{{Hamers} {et~al.}(2015){Hamers}, {Perets}, {Antonini}, \& {Portegies
  Zwart}}]{Hamersetal2015}
{Hamers}, A.~S., {Perets}, H.~B., {Antonini}, F., \& {Portegies Zwart}, S.~F.
  2015, Monthly Notices of the Royal Astronomical Society, 449, 4221

\bibitem[{{Hansen}(1855)}]{Hansen1855}
{Hansen}, P.~A. 1855, Abhandlungen der Koniglich Sachsischen Gesellschaft der
  Wissenschaften, 2, 183, {English translation by J.C. Van der Ha, ESA/ESOC,
  Darmstadt, Germany, 1977}

\bibitem[{{Hansen}(1857)}]{Hansen1857}
{Hansen}, P.~A. 1857, Abhandlungen der Koniglich Sachsischen Gesellschaft der
  Wissenschaften, 5, 41

\bibitem[{{Henrard}(1974)}]{Henrard1974}
{Henrard}, J. 1974, Celestial Mechanics, 10, 437

\bibitem[{{Herrick}(1948)}]{Herrick1948}
{Herrick}, S. 1948, Publications of the Astronomical Society of the Pacific,
  60, 321

\bibitem[{Hintz(2008)}]{Hintz2008}
Hintz, G. 2008, Journal of Guidance, Control, and Dynamics, 31, 785

\bibitem[{{Hori}(1966)}]{Hori1966}
{Hori}, G.-i. 1966, Publications of the Astronomical Society of Japan, 18, 287

\bibitem[{{Katz} {et~al.}(2011){Katz}, {Dong}, \&
  {Malhotra}}]{KatzDongMalhotra2011}
{Katz}, B., {Dong}, S., \& {Malhotra}, R. 2011, Physical Review Letters, 107,
  181101

\bibitem[{{Kaula}(1962)}]{Kaula1962}
{Kaula}, W.~M. 1962, The Astronomical Journal, 67, 300

\bibitem[{{Kaula}(1966)}]{Kaula1966}
{Kaula}, W.~M. 1966, {Theory of satellite geodesy. Applications of satellites
  to geodesy} (Waltham, Massachusetts: Blaisdell)

\bibitem[{{Kelly}(1989)}]{Kelly1989}
{Kelly}, T.~S. 1989, Celestial Mechanics and Dynamical Astronomy, 46, 19

\bibitem[{{Lane}(1989)}]{Lane1989}
{Lane}, M.~T. 1989, Celestial Mechanics and Dynamical Astronomy, 46, 287

\bibitem[{{Lara}(2016)}]{LaraTossa2016}
{Lara}, M. 2016, in Astrophysics and Space Science Proceedings, Vol.~44,
  Astrodynamics Network AstroNet-II: The Final Conference, ed. G.~{G{\'o}mez}
  \& J.~{Masdemont} (Cham: Springer), 151--166

\bibitem[{{Lara}(2017)}]{Lara2017if}
{Lara}, M. 2017, Celestial Mechanics and Dynamical Astronomy, 129, 137

\bibitem[{Lara {et~al.}(2016)Lara, San-Juan, \&
  Hautesserres}]{LaraSanJuanHautesserresCNES2016}
Lara, M., San-Juan, J., \& Hautesserres, D. 2016, {Semi-analytical propagator
  of high eccentricity orbits}, Technical Report R-S15/BS-0005-024, Centre
  National d'\'Etudes Spatiales, 18, avenue Edouard Belin - 31401 Toulouse
  Cedex 9, France

\bibitem[{{Lara} {et~al.}(2018){Lara}, {San-Juan}, \&
  {Hautesserres}}]{LaraSanJuanHautesserres2018}
{Lara}, M., {San-Juan}, J.~F., \& {Hautesserres}, D. 2018, CEAS Space Journal,
  10, 3

\bibitem[{{Lara} {et~al.}(2012){Lara}, {San-Juan}, {L{\'o}pez}, \&
  {Cefola}}]{LaraSanJuanLopezCefola2012}
{Lara}, M., {San-Juan}, J.~F., {L{\'o}pez}, L.~M., \& {Cefola}, P.~J. 2012,
  Celestial Mechanics and Dynamical Astronomy, 113, 435

\bibitem[{{Laskar} \& {Bou\'e}(2010)}]{LaskarBoue2010}
{Laskar}, J. \& {Bou\'e}, G. 2010, {Astronomy and Astrophysics}, 522, A60

\bibitem[{{Laskar} \& {Robutel}(2001)}]{LaskarRobutel2001}
{Laskar}, J. \& {Robutel}, P. 2001, Celestial Mechanics and Dynamical
  Astronomy, 80, 39

\bibitem[{{Lee} \& {Peale}(2003)}]{LeePeale2003}
{Lee}, M.~H. \& {Peale}, S.~J. 2003, The Astrophysical Journal, 592, 1201

\bibitem[{{Libert} \& {Sansottera}(2013)}]{LibertSansottera2013}
{Libert}, A.~S. \& {Sansottera}, M. 2013, Celestial Mechanics and Dynamical
  Astronomy, 117, 149

\bibitem[{{Mardling}(2013)}]{Mardling2013}
{Mardling}, R.~A. 2013, Monthly Notices of the Royal Astronomical Society, 435,
  2187

\bibitem[{{Meyer} \& {Hall}(1992)}]{MeyerHall1992}
{Meyer}, K.~R. \& {Hall}, G.~R. 1992, {Introduction to Hamiltonian Dynamical
  Systems and the N-Body Problem} (New York: Springer)

\bibitem[{{Migaszewski} \&
  {Go{\'z}dziewski}(2008)}]{MigaszewskiGozdziewski2008}
{Migaszewski}, C. \& {Go{\'z}dziewski}, K. 2008, Monthly Notices of the Royal
  Astronomical Society, 388, 789

\bibitem[{{Mignard} \& {Henon}(1984)}]{MignardHenon1984}
{Mignard}, F. \& {Henon}, M. 1984, Celestial Mechanics, 33, 239

\bibitem[{Milankovitch(1941)}]{Milankovitch1941}
Milankovitch, M. 1941, {Kanon der Erdbestrahlung und seine Anwendung auf das
  Eiszeitenproblem}, Mechanics of Space Flight (Belgrade: K\"oniglich Serbische
  Akademie), {English translation: Canon of Insolation and the Ice-age Problem.
  Israel Program for Scientific Translations, Jerusalem, 1969}

\bibitem[{Musen(1961)}]{Musen1961}
Musen, P. 1961, Journal of Geophysical Research, 66, 2797

\bibitem[{{Musen}(1963)}]{Musen1963JGR}
{Musen}, P. 1963, Journal of Geophysical Research, 68, 6255

\bibitem[{{Palaci{\'a}n} {et~al.}(2017){Palaci{\'a}n}, {Vanegas}, \&
  {Yanguas}}]{PalacianVanegasYanguas2017}
{Palaci{\'a}n}, J.~F., {Vanegas}, J., \& {Yanguas}, P. 2017, Astrophysics and
  Space Science, 362, 215

\bibitem[{{Richter} \& {Keller}(1995)}]{RichterKeller1995}
{Richter}, K. \& {Keller}, H.~U. 1995, Icarus, 114, 355

\bibitem[{{Rosengren} \& {Scheeres}(2013)}]{RosengrenScheeres2013}
{Rosengren}, A. \& {Scheeres}, D. 2013, Advances in Space Research, 52, 1545

\bibitem[{{Rosengren} \& {Scheeres}(2014)}]{RosengrenScheeres2014}
{Rosengren}, A.~J. \& {Scheeres}, D.~J. 2014, Celestial Mechanics and Dynamical
  Astronomy, 118, 197

\bibitem[{{Roy} \& {Moran}(1973)}]{RoyMoran1973}
{Roy}, A.~E. \& {Moran}, P.~E. 1973, Celestial Mechanics, 7, 236

\bibitem[{{Sansottera} \& {Libert}(2019)}]{SansotteraLibert2019}
{Sansottera}, M. \& {Libert}, A.~S. 2019, Celestial Mechanics and Dynamical
  Astronomy, 131, 38

\bibitem[{{Will}(2017)}]{Will2017}
{Will}, C.~M. 2017, Physical Review D, 96, 023017

\end{thebibliography}

\end{document}